\documentclass[letterpaper, 10 pt, conference,final]{ieeeconf}  

\IEEEoverridecommandlockouts                              

\usepackage{amsmath} 

\title{\Huge 
Network Flow Routing under Strategic Link Disruptions
}
\author{\authorblockN{Mathieu Dahan}
\authorblockA{Center for Computational Engineering (CCE)\\
Massachusetts Institute of Technology\\
Cambridge, MA 02139\\
Email: mdahan@mit.edu}
\and
\authorblockN{Saurabh Amin}
\authorblockA{Department of Civil \& Environmental Engineering\\
Massachusetts Institute of Technology\\
Cambridge, MA 02139\\
Email: amins@mit.edu}}

\usepackage{amssymb}
\usepackage{stmaryrd}
\usepackage{tikz}
\usetikzlibrary{arrows}
\usepackage{color}
\usepackage{dsfont}
\usepackage{xspace}
\usepackage{subcaption}

\begin{document}

\newtheorem{lemma}{\textbf{Lemma}}
\newtheorem{proposition}{\textbf{Proposition}}
\newtheorem{definition}{\textbf{Definition}}
\newtheorem{theorem}{\textbf{Theorem}}
\newtheorem{remark}{\textbf{Remark}}
\newtheorem{assumption}{\textbf{Assumption}}
\newtheorem{corollary}{\textbf{Corollary}}
\newtheorem{example}{\textbf{Example}}

\newcommand{\defender}{\textbf{P1}\xspace}
\newcommand{\attacker}{\textbf{P2}\xspace}

\newcommand{\supp}{\operatorname{supp}}
\newcommand{\Val}{\operatorname{F}}
\newcommand{\Costop}{\operatorname{C_2}}
\newcommand{\Costflow}{\operatorname{C_1}}
\newcommand{\flow}{x}
\newcommand{\att}{\mu}

\newcommand{\Value}[1]{\Val\left(#1\right)}
\newcommand{\Eff}[2]{\Val\left({#1}^{#2}\right)}
\newcommand{\Effp}[2]{\Val\left((#1)^{#2}\right)}
\newcommand{\Loss}[2]{\Val\left(#1 - {#1}^{#2}\right)}
\newcommand{\Lossp}[2]{\Val\left(#1 - (#1)^{#2}\right)}
\newcommand{\Cost}[1]{\Costop\left(#1\right)}
\newcommand{\Costf}[1]{\Costflow(#1)}
\newcommand{\Exponebis}[2]{\mathbb{E}_{#1}\left[#2\right]}
\newcommand{\Exptwobis}[2]{\mathbb{E}_{#1}\left[#2\right]}
\newcommand{\Exps}[3]{\mathbb{E}_{(#1,#2)}\left[#3\right]}
\newcommand{\Expone}[2][\empty]{\mathbb{E}_{{\sigma}^{#1}}\left[#2 \right]}
\newcommand{\Exptwo}[2][\empty]{\mathbb{E}_{{\sigma}^{#1}}\left[#2 \right]}
\newcommand{\Expboth}[3]{\mathbb{E}_{\sigma^{#1}}\left[#3\right]}

\maketitle
\thispagestyle{empty}
\pagestyle{empty}

\begin{abstract}

This paper considers a 2-player strategic game for network routing under link disruptions. Player $1$ (defender) routes flow through a network to maximize her value of effective flow while facing transportation costs. Player $2$ (attacker) simultaneously disrupts one or more links to maximize her value of lost flow but also faces cost of disrupting links. This game is strategically equivalent to a zero-sum game. Linear programming duality and the max-flow min-cut theorem are applied to obtain properties that are satisfied in any mixed Nash equilibrium. In any equilibrium, both players achieve identical payoffs. While the defender's expected transportation cost decreases in attacker's marginal value of lost flow, the attacker's expected cost of attack increases in defender's marginal value of effective flow. Interestingly, the expected amount of effective flow decreases in both these parameters. These results can be viewed as a generalization of the classical max-flow with minimum transportation cost problem to adversarial environments. 

\end{abstract}


\section{Introduction.}\label{intro}

We study a two-player strategic game over a directed network in which player~$1$ (defender or agency) chooses a flow to be routed from a source node to a destination node, and player~$2$ (attacker or interdictor) chooses to disrupt one or more edges. The payoff structures in this game are motivated by the previous formulations in both network interdiction problems~\cite{Wood19931,Morton_98,BertsimasNS13} and attacker-defender games~\cite{DBLP:conf/gamesec/GueyeWA10,Szeto,Baykal-GursoyDPG14}. In our model, we account the value of the effective flow and the transportation cost for the defender, and the value of the lost flow and the cost of attack for the attacker.  
Specifically, the player~$1$'s payoff increases in the amount of effective flow that reached the destination node, but decreases with the cost of transporting the initial flow chosen by her. The player~$2$'s payoff increases in the amount of lost flow as a result of an attack and decreases with the size of attack. This two player game is strictly equivalent to a strictly competitive game (SCG).   

We characterize the Nash equilibria for this game for a special class of networks. We show that our game retains some of the nice properties of SCGs. For example, we prove that each player has a unique payoff value in all equilibria. We show that every Nash equilibrium gives a minimaximizing strategy for each player, but the converse is not true. Interestingly, there is a unique maximinimizing strategy but it is not a Nash strategy. Using a combination of game-theoretic ideas and duality results such as the max-flow min-cut theorem, we are able to compute the values of effective and lost flow and the costs of transportation and attack in terms of the parameters of the game. We also relate the structure of mixed strategy Nash equilibrium of the game to the solutions of the maximum flow minimum transportation cost problem and to the minimum cuts.


%

Network interdiction problems have already been widely studied, but our focus is to extend these formulations to simultaneous game settings. Related to our approach is the article by Washburn and Wood~\cite{Washburn1995}. The authors model a sequential game, where the defender (leader) chooses one $s-t$ path and then the interdictor (follower) inspects one arc. The objective of this sequential game model is to maximize the probability with which the operator is detected by the interdictor. Our game differs from the model in~\cite{Washburn1995} in that we model a simultaneous game, which captures each player's strategic uncertainty about her opponent. Secondly, we allow both players to have a much larger set of actions (feasible flow that may contain many $s-t$ paths and loops, and attacks that can disrupts several edges at the same time). Finally, we account for the attacker's cost of attack as well as the defender's cost of transporting flow through the network. 

Another relevant line of work is by Bertsimas et al.~\cite{BertsimasNO13}. In this sequential game, the operator first chooses a feasible flow, and then the interdictor disrupts a fixed number of edges. The goal is to minimize the largest amount of flow that reaches the destination node. The authors  consider two different models for the disruption: an arc-based formulation where the flow can be rerouted when there is an attack, and a path-based formulation where the flow carried by a disrupted edge is lost. Our formulation is related to the path-based formulation: since we model a simultaneous game, it is reasonable to assume that the flow through disrupted edges is lost and cannot be re-routed. Although, in~\cite{BertsimasNO13}, the interdictor can disrupt several edges at the same time, she must always disrupt the same number of edges for every action, which is still a restriction of the set of actions we considered for our interdictor.

The work of Gueye et al.~\cite{DBLP:conf/gamesec/GueyeWA10} (also see~\cite{DBLP:conf/gamesec/LaszkaG13}) is one of the first works to model a simultaneous attacker-defender game, where the defender (operator) chooses how to send a feasible flow that satisfies the supply and demand constraints (these are exogenous parameters), and the attacker disrupts a single edge of the network. Once again, the utility of the operator only takes into account the amount of flow that reaches the destination node. Similar to~\cite{DBLP:conf/gamesec/GueyeWA10}, our model also considers that the attacker's payoff depends on the amount of lost flow and on the cost of attack. However, in our model, the cost of attack is proportional to the capacities of the edges whereas Gueye et al. only consider a constant cost of attack. Another notable difference is that ~\cite{DBLP:conf/gamesec/GueyeWA10} only considers an uncapacitated graph with given supplies and demands, and we consider a capacitated graph with no constraint on the supplies and demands.

Finally, our model is closely related to the work of Hong and Wooders~\cite{RePEc:van}. The authors model a two player simultaneous game in which an operator chooses a feasible flow  and an interdictor disrupts edges of the network, preventing the flow from reaching the destination node. The operator's utility accounts for the amount of flow that reaches the terminal node and the corresponding cost of transportation. However, in~\cite{RePEc:van}, the operator's utility is assumed to increase with the cost of attack. This is a central point of departure between their work and ours. Specifically, in our model, the interdictor's utility decreases with the cost of attack. While the results in~~\cite{RePEc:van} are based on a max-flow and a min-cut of the graph, we relate Nash equlibria of our game with the solutions of the max-flow with minimum transportation cost problem and with min-cuts.

The rest of the paper is organized as follows: In Section~\ref{sec:problem} we discuss the main assumptions and present our game model. Section~\ref{sec:NE} presents our main results on the characterization of Nash equilibria of the game. The implications of relaxing some of the modeling assumptions are discussed in~Section~\ref{sec:relax}.
\section{Problem}\label{sec:problem}

\subsection{Preliminaries}
\label{ss:prelim}

Consider a capacitated directed graph $\mathcal{G} = (\mathcal{V},\mathcal{E})$ where $\mathcal{V}$ (resp. $\mathcal{E})$ represents the set of nodes (resp. the set of edges) of $\mathcal{G}$. For each edge $(i,j) \in \mathcal{E}$, let $c_{ij} \in \mathbb{R}^+$ denote its capacity, which is the maximum amount of commodity that can pass through $(i,j)$. Let $s\in \mathcal{V}$ denote a source node and $t\in \mathcal{V}$ a destination node. The flow can only enter the network from $s$ and leave from $t$, and there is no demand or supply at the nodes that are different from $s$ and $t$. A flow, denoted $\flow$, is a function from $\mathcal{E}$ to $\mathbb{R}^+$ that assigns to each edge the amount of commodity that goes through it. For notational simplicity, let   $\flow_{ij} := \flow((i,j))$ denote the flow through edge $(i,j)$. Let $\mathcal{F}$ denote the set of feasible flows, where a flow $\flow$ is said to be feasible if it satisfies flow conservation at each node and if the flow through each edge does not exceed its capacity: 
\begin{align*}
\forall i \in \mathcal{V}\backslash \{s,t\}, & \ \displaystyle\sum_{(j,i) \in \mathcal{E}} \flow_{ji} = \sum_{(i,j) \in \mathcal{E}} \flow_{ij}, \\ 
\forall (i,j) \in \mathcal{E}, & \ 0 \leq \flow_{ij} \leq c_{ij}.
\end{align*} 
Let $\Lambda$ denote the set containing all the loops and $s-t$ paths (i.e., a path that starts from $s$ and ends at $t$) of the network, and $\flow_{\lambda}$ the quantity of flow of ${\flow}$ sent through $\lambda \in \Lambda$. Then the edge flows and path/loop flows satisfy: 
\begin{align}
\label{eq:pathstoedges}
\forall (i,j) \in \mathcal{E}, \ \flow_{ij} = \sum_{\{\lambda \in \Lambda \, | \, (i,j) \in \lambda\}}\flow_{\lambda}
\end{align}
An \emph{$s-t$ cut} is a partition $\{S,T\}$ of the set of nodes $\mathcal{V}$, such that $s \in S$ and $t \in T$. We define the \emph{cut-set} of $\{S,T\}$ as $E(\{S,T\}) = \left\{(i,j) \in \mathcal{E} \, |\, i \in S, \, j \notin S \right\}$ and the \emph{capacity} of $\{S,T\}$ as $C(\{S,T\}) = \sum_{(i,j) \in E(\{S,T\})} c_{ij}$. Let us recall the \emph{max-flow problem}:
\begin{align*}
\begin{array}{rll}\text{maximize} & \displaystyle\Value{\flow} &  \\ \text{subject to} & \displaystyle\sum_{(j,i) \in \mathcal{E}} \flow_{ji} = \sum_{(i,j) \in \mathcal{E}} \flow_{ij}, & \forall i \in \mathcal{V}\backslash \{s,t\}\\ & 0 \leq \flow_{ij} \leq c_{ij}, & \forall (i,j) \in \mathcal{E}\end{array}
\end{align*}
where $\Value{\flow} =\sum_{\left\{ i \in \mathcal{V}\, | \, (i,t) \in \mathcal{E} \right\}}\flow_{it} $ denotes the amount of flow passing from the source $s$ to the sink $t$. The well-known \emph{max-flow min-cut theorem} by Ford and Fulkerson \cite{Ford-Fulkerson_algo} states that the optimal value of the maximum flow problem is equal to the minimum capacity over all {$s-t$ cuts}. We use $\Theta$ to denote the optimal value of the max-flow problem. 

We also state the \emph{maximum flow minimum transportation cost problem} by Edmonds and Karp \cite{Edmonds:1972aa} in which the goal is to find a maximum flow with minimum transportation cost:
\begin{align*}
\begin{array}{lrll}(\mathcal{P})& \text{minimize} & \displaystyle\sum_{(i,j) \in \mathcal{E}}b_{ij}\flow_{ij} &  \\ & \text{subject to} & \flow \in \mathcal{F} & \\& &\displaystyle \Value{\flow} \geq \Value{\flow^{\prime}}, & \forall \flow^{\prime} \in \mathcal{F}\end{array}
\end{align*}
where for every edge $(i,j) \in \mathcal{E}$, $b_{ij}\in\mathbb{R}^+$ denotes the cost of transporting one unit of flow through $(i,j)$. We denote the set of optimal solutions of problem $(\mathcal{P})$ by $\Omega$.

\subsection{Model}
\label{ss:model}

We focus on a simultaneous two-player strategic game $\Gamma :=\langle\{1,2\},(\mathcal{F},\mathcal{A}),(u_1,u_2)\rangle$ defined as follows: player~$1$ (\defender) is the defender (agency) who chooses to route a flow $\flow\in \mathcal{F}$ through the network, and player~$2$ (\attacker) is the attacker (interdictor) who chooses an attack $\att$ to disrupt a subset of edges of graph $\mathcal{G}$; see~\eqref{def_attack}. The set of actions for \defender (resp. \attacker) is given by $\mathcal{F}$ (resp. $\mathcal{A} := \{0,1\}^{\mathcal{E}}$). 

An attack $\att$ is a function from $\mathcal{E}$ to $\{0,1\}$ defined as follows:
\begin{align}
{\att}_{ij} := \att((i,j)) =\left\{ \begin{array}{cl} 1 & \text{if} \ (i,j) \ \text{is disrupted}, \\ 0 & \text{otherwise.} \end{array} \right.
\label{def_attack}
\end{align}
For our simultaneous game, it is reasonable to assume that after an edge is disrupted (or attacked) by \attacker, the flow that was supposed to cross this edge (if there were no attack) is lost and not re-routed.\footnote{We do not consider attacks that can only result in partially disrupted edges and might still permit some flow to pass through the attacked edges.} Thus, the effective flow, denoted ${\flow}^{\att}$, when the flow ${\flow}$ is chosen by \defender and the attack $\att$ is chosen by \attacker can be expressed as follows:
\begin{align*}
\forall (i,j) \in \mathcal{E}, \ \flow_{ij}^{\att} = \sum_{\lambda \in \Lambda_{ij}^{\att}}\flow_{\lambda},
\end{align*}
with $\Lambda_{ij}^{\att} := \left\{\lambda \in \Lambda \, | \, (i,j) \in \lambda \ \text{and} \ \forall (i^{\prime},j^{\prime}) \in \lambda, {\att}_{i^{\prime}j^{\prime}} = 0 \right\}$. That is, the effective flow through an edge is the sum of all the initial path flows that go through that edge and that do not contain any attacked edge. 
\vspace{0.1cm}
\begin{example}
Consider the flow chosen by \defender in Fig.~\ref{fig:Initial flow}. If \attacker chooses to disrupt edges $(1,t)$ and $(s,2)$, the unit flows through paths $\{s,1,t\}$ and $\{s,2,t\}$ are lost. The resulting effective flow is shown in Fig~\ref{fig:Eff}.

\begin{figure}[htbp]
    \centering
    \begin{subfigure}[c]{0.23\textwidth}
        \centering
        \begin{tikzpicture}[->,>=stealth',shorten >=1pt,auto,x=1.6cm, y=1cm,
  thick,main node/.style={circle,draw}, main2 node/.style={},flow_a/.style ={blue!100}]
\tikzstyle{edge} = [draw,thick,->]
\tikzstyle{cut} = [draw,very thick,-]
\tikzstyle{flow} = [draw,line width = 1pt,->,blue!100]
\small
	\node[main node] (s) at (0,1) {s};
	\node[main node] (1) at (1,2) {1};
	\node[main node] (2) at (1,0) {2};
	\node[main node] (t) at (2,1) {t};

	\path[flow]
	(s) edge node{} (1)	
	(1) edge node{}(2)
	(2) edge node{}(t)
	(s) edge node{}(2)
	(1) edge node{}(t);

\node[main2 node] (e) at (0.2,1.8) {\scriptsize$\begin{array}{c}\textcolor{blue}{\flow_{s1} = 2} \\ \textcolor{red}{\att_{s1} = 0}\end{array}$};

\node[main2 node] (e) at (1.35,1) {\scriptsize$\begin{array}{c}\textcolor{blue}{\flow_{12} = 1} \\ \textcolor{red}{\att_{12} = 0}\end{array}$};

\node[main2 node] (e) at (1.85,0.25) {\scriptsize$\begin{array}{c}\textcolor{blue}{\flow_{2t} = 2} \\ \textcolor{red}{\att_{2t} = 0}\end{array}$};

\node[main2 node] (e) at (0.2,0.2) {\scriptsize$\begin{array}{c}\textcolor{blue}{\flow_{s2} = 1} \\ \textcolor{red}{\att_{s2} = 1}\end{array}$};

\node[main2 node] (e) at (1.85,1.85) {\scriptsize$\begin{array}{c}\textcolor{blue}{\flow_{1t} = 1} \\ \textcolor{red}{\att_{1t} = 1}\end{array}$};

\draw[very thick, red, dashed,cut] 
    (0.5,0.2).. controls +(right:0cm) and +(down:0cm) ..(0.75,0.6);
    
      \draw[very thick, red, dashed,cut] 
    (1.5,1.8).. controls +(right:0cm) and +(down:0cm) ..(1.25,1.4);
\normalsize
\end{tikzpicture}
        \caption{Initial flow and attack.}
	\label{fig:Initial flow}
    \end{subfigure}
    ~ 
    \begin{subfigure}[c]{0.23\textwidth}
        \centering
        \begin{tikzpicture}[->,>=stealth',shorten >=1pt,auto,x=1.6cm, y=1cm,
  thick,main node/.style={circle,draw},flow_a/.style ={blue!100},node bis/.style={}]
\tikzstyle{edge} = [draw,thick,dashed,->]
\tikzstyle{cut} = [draw,very thick,-]
\tikzstyle{flow} = [draw,line width = 1pt,->,blue!100]
\small
	\node[main node] (s) at (0,1) {s};
	\node[main node] (1) at (1,2) {1};
	\node[main node] (2) at (1,0) {2};
	\node[main node] (t) at (2,1) {t};
	
	\path[edge]
	(s) edge node[below left]{\scriptsize$\flow^{\att}_{s2} = 0$}(2)
	(1) edge node{\scriptsize$\flow^{\att}_{1t} = 0$}(t);
	
	\path[flow]
	(s) edge node{\scriptsize$\flow^{\att}_{s1} = 1$} (1)
	(1) edge node{\scriptsize$\flow^{\att}_{12} = 1$}(2)
	(2) edge node[below right]{\scriptsize$\flow^{\att}_{2t} = 1$}(t);

  \draw[very thick, red,cut] 
    (0.5,0.2).. controls +(right:0cm) and +(down:0cm) ..(0.75,0.6);
    
      \draw[very thick, red,cut] 
    (1.5,1.8).. controls +(right:0cm) and +(down:0cm) ..(1.25,1.4);
    
    \normalsize
\end{tikzpicture}
        \caption{Resulting effective flow.}
	\label{fig:Eff}
    \end{subfigure}
    \caption{Example graph.}
    \label{fig:Example}
\end{figure}
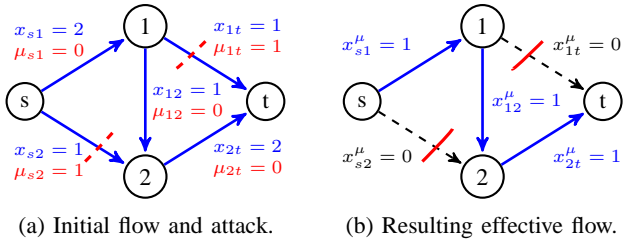
\end{example}
\vspace{0.1cm}

The payoff of \defender is defined as follows:
\begin{align}\label{eq:p1payoff}
u_1({\flow},{\att}) = \underset{\text{value of effective flow}}{\underbrace{p_1 \Eff{\flow}{\att}}} - \underset{\text{transportation cost}}{\underbrace{\Costf{\flow}}} ,
\end{align}
where $p_1\in\mathbb{R}^+$ is the marginal value of the flow assessed by \defender and $\Costf{\flow}:=\sum_{(i,j) \in \mathcal{E}} b_{ij}\flow_{ij}$ is the cost of transporting the initial flow~$\flow$. 
Thus, when one additional unit of flow reaches $t$, \defender's payoff increases by $p_1$ and at the same time decreases by its transportation cost. 


The payoff of \attacker is defined as follows:
\begin{align}\label{eq:p2payoff}
u_2({\flow},{\att}) =\underset{\text{value of lost flow}}{\underbrace{p_2 \Loss{\flow}{\att}}} - \underset{\text{cost of attack}}{\underbrace{\Cost{\att}}},
\end{align}
where $p_2\in\mathbb{R}^+$ is the marginal value of the lost flow to \attacker, (in general, $p_1\neq p_2$) and $\Cost{\att}:=\sum_{(i,j) \in \mathcal{E}} c_{ij}\att_{ij}$ is the cost of the attack $\att$. 
Thus, if the disruption of an edge induces the loss of one unit of flow, the payoff of \attacker increases by $p_2$, and at the same time decreases by the disruption cost, equal to the edge capacity.


We allow both players to randomize over their set of pure actions. Let $\displaystyle \Delta(\mathcal{F})$ and $\displaystyle \Delta(\mathcal{A})$ denote the mixed extensions of \defender and \attacker pure strategies, respectively, i.e.: 
\small
\begin{align*}
\displaystyle \Delta(\mathcal{F}) &= \left\{\sigma^1 \in [0,1]^{\mathcal{F}}\ \Big| \ \sum_{{\flow} \in \mathcal{F}} \sigma^1(\flow) = 1 \right\}, \\ \displaystyle \Delta(\mathcal{A}) &= \left\{\sigma^2 \in [0,1]^{\mathcal{A}}\ \Big| \ \sum_{{\att} \in \mathcal{A}} \sigma^2({\att}) = 1 \right\}.\end{align*} 
\normalsize
For notational simplicity, we define $\sigma^1_{\flow} := \sigma^1({\flow})$ and $\sigma^2_{\att} := \sigma^2({\att})$. Given any function $\varphi : \mathcal{F} \times \mathcal{A} \longrightarrow \mathbb{R}$ and any strategy profile $\sigma = (\sigma^1,\sigma^2) \in \Delta(\mathcal{F}) \times \Delta(\mathcal{A})$, we denote $\Expboth{}{}{\varphi(\flow,\att)} := \sum_{\flow \in \mathcal{F}} \sigma^1_{\flow} \sum_{\att \in \mathcal{A}} \sigma^2_{\att} \,\varphi(\flow,\att)$ the expectation of $\varphi$ with respect to $\sigma$. Then, the respective expected payoffs can be expressed as:
\normalsize
\begin{align}
U_1(\sigma^1,\sigma^2) & =p_1 \Expboth{}{}{\Eff{\flow}{\att}} - \Expone{\Costf{\flow}} \label{payoff1}\\
U_2(\sigma^1,\sigma^2) &= p_2 \Expone{\Loss{\flow}{\att}} - \Exptwo{\Cost{\att}}\label{payoff2}
\end{align}
\normalsize

A mixed strategy profile $({\sigma^1}^\ast,{\sigma^2}^\ast) \in \Delta(\mathcal{F}) \times \Delta(\mathcal{A})$ is a Nash Equilibrium (NE) if:
\begin{align}
\forall {\sigma^1} \in \Delta(\mathcal{F}), \ U_1({\sigma^1}^*,{\sigma^2}^*) \geq U_1({\sigma^1},{\sigma^2}^*) \label{best1},\\
\forall {\sigma^2} \in \Delta(\mathcal{A}), \ U_2({\sigma^1}^*,{\sigma^2}^*) \geq U_2({\sigma^1}^*,{\sigma^2}). 
\label{best2}
\end{align}

Equivalently, at NE, ${\sigma^1}$ (resp. ${\sigma^2}$) is a best response to $\sigma^2$ (resp. $\sigma^1$).
The \emph{support} of $\sigma^1$ (resp. $\sigma^2$) is $\supp(\sigma^1) = \{\flow \in \mathcal{F} \ | \ \sigma^1_{\flow} >0 \}$ (resp. $\supp(\sigma^2) = \{\att \in \mathcal{A} \ | \ \sigma^2_{\att} >0 \}$). 
%
%

Our objective is to characterize the set of NE of $\Gamma $, denoted $\mathcal{S}_{\Gamma}$, under the following assumption: 
\vspace{0.1cm}
\begin{assumption}
\textit{Let $ \alpha := \min_{\lambda \in \Lambda_{path}} \sum_{(i,j) \in \lambda} b_{ij}$. There exists an optimal solution of $(\mathcal{P})$, denoted $\flow^{\ast}$ that takes $s-t$ paths with marginal transportation cost equal to $\alpha$, i.e., }
\begin{align*}
\exists\, \flow^{\ast} \in \Omega \ | \ \forall \lambda \in \Lambda_{path}, \ \flow_{\lambda} >0 \Longrightarrow \sum_{(i,j) \in \lambda} b_{ij} = \alpha, 
\end{align*}
\textit{where $\Lambda_{path}$ is the set containing all the $s-t$ paths of the network.}
\label{big_assumption}
\end{assumption}
\vspace{0.1cm}

This assumption implies that if $\flow^{\ast}\in \Omega$ denotes a max-flow with minimum transportation cost, the cost of transporting a unit flow through each $s-t$ path taken by $\flow^{\ast}$ is identically equal to $\alpha$. In addition, every other path in the network cannot have a smaller marginal transportation cost. Note that the case when every $s-t$ has an identical marginal transportation cost is a special case of this assumption. We illustrate Assumption~\ref{big_assumption} with the following example:
\vspace{0.1cm}
\begin{example} Consider the network flow problem in~Fig.~\ref{fig:Max-flow min Cost}. There is a unique maximum flow with minimum transportation cost $\flow^{\ast}$ which carries $1$ unit of flow through paths $\{s,2,4,t\}$, $\{s,2,3,t\}$ and $\{s,1,t\}$. Thus, the total amount of flow is equal to $3$ units. In this network, $\alpha = 3$, and each path taken by $\flow^{\ast}$ has a marginal transportation cost equal to $3$ so the cost of transporting $\flow^{\ast}$ is equal to 9. The remaining paths that are not taken by $\flow^{\ast}$ are $\{s,4,t\}$ with a transportation cost equal to 4, and $\{s,1,3,t\}$ with a transportation cost equal to 3. Thus, Assumption~\ref{big_assumption} is satisfied.

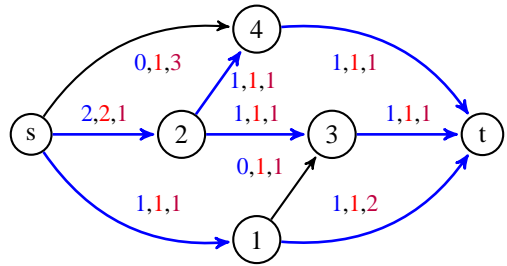
\begin{figure}[htbp]
\centering
\begin{tikzpicture}[->,>=stealth',shorten >=1pt,auto,x=2cm, y=1.4cm,
  thick,main node/.style={circle,draw},flow_a/.style ={blue!100}]
\tikzstyle{edge} = [draw,thick,->]
\tikzstyle{cut} = [draw,very thick,-]
\tikzstyle{flow} = [draw,line width = 1pt,->,blue!100]
	\node[main node] (s) at (0,1) {s};
	\node[main node] (1) at (1.5,0) {1};
	\node[main node] (2) at (1,1) {2};
	\node[main node] (3) at (2,1) {3};
	\node[main node] (4) at (1.5,2) {4};
	\node[main node] (t) at (3,1) {t};
	
	\path[edge]
	(s) edge [bend left]node[below right] {\small\textcolor{blue}{0},\textcolor{red}{1},\textcolor{purple}{3}} (4)
	(1) edge node  {\small\textcolor{blue}{0},\textcolor{red}{1},\textcolor{purple}{1}} (3);

	\path[flow,every node/.style={black!100}]
	(s) edge[bend right] node[above right] {\small\textcolor{blue}{1},\textcolor{red}{1},\textcolor{purple}{1}} (1)
	(s) edge node{\small\textcolor{blue}{2},\textcolor{red}{2},\textcolor{purple}{1}} (2)
	(2) edge node[right]{\small\textcolor{blue}{1},\textcolor{red}{1},\textcolor{purple}{1}} (4)
	(2) edge node{\small\textcolor{blue}{1},\textcolor{red}{1},\textcolor{purple}{1}} (3)
	(3) edge node{\small\textcolor{blue}{1},\textcolor{red}{1},\textcolor{purple}{1}} (t)
	(4) edge [bend left]node[below left]{\small\textcolor{blue}{1},\textcolor{red}{1},\textcolor{purple}{1}} (t)
	(1) edge [bend right]node[above left]{\small\textcolor{blue}{1},\textcolor{red}{1},\textcolor{purple}{2}} (t);
	
	
\end{tikzpicture}
\caption{Max-flow with minimum transportation cost (drawn in blue). The labels of each edge correspond to the flow it carries (blue), its capacity (red) and its transportation cost (purple).}
\label{fig:Max-flow min Cost}
\end{figure}
\end{example}
\vspace{0.1cm}

Admittedly, on one hand, Assumption~\ref{big_assumption} restricts the class of networks on which the routing game~$\Gamma$ is played. On the other hand, it permits us to obtain strong results on the equilibrium structure.
%

Note that $\Gamma$ can be easily extended to networks with multiple sources and multiple destination nodes. Given such a network, one needs to add an extra source (resp. destination) node and connect it to every existing source (resp. destination) node with an uncapacitated edge of transportation cost equal to 0. This modification gives a new network with single source and single destination. The outcome of the game defined for the original network remains the same as that of the game defined for the new network.

\section{Characterization of NE of $\Gamma$}\label{sec:NE}



In this section, we study the properties of the set of NE $\mathcal{S}_{\Gamma}$. First, we argue that $\Gamma$ is \emph{not} an SCG. Recall that a two-person game is said to be strictly competitive if, when both players change their mixed strategies, either the expected payoffs remain the same or one of the expected payoffs strictly increases and the other strictly decreases. More precisely, $\Gamma$ can be shown to be not an SCG by arguing that $u_1$ is not an affine variant of $-u_2$ in the sense of Adler et al.~\cite{DBLP:conf/wine/AdlerDP09}. Although $\Gamma$ is not an SCG, we can show that $\Gamma$ is strategically equivalent to a zero-sum game. The strategic equivalence between $\Gamma$ and a zero-sum game implies that $\mathcal{S}_{\Gamma}$ is a convex set.


First, we parametrically solve the game $\Gamma$ and we present NE whose support involves optimal solutions of $(\mathcal{P})$ and min-cuts.

Secondly, we focus on the region $p_1>\alpha$, $p_2>1$ which only admits mixed NE and we present our main theorem that provides analytical expressions of certain quantities of interest at any NE. Specifically, we prove that each player has a unique payoff value in all $\sigma^\ast \in \mathcal{S}_{\Gamma}$ and we characterize the value of effective (resp. lost) flow and the cost of transportation (resp. cost of attack) in terms of the parameters of $\Gamma$: $p_1$, $p_2$, and $\alpha$.

Third, we relate the mixed strategy NE of $\Gamma$ to the solutions of the max-flow minimum transportation cost problem ($\mathcal P$), and also to the minimum cuts. Finally, we can show that for every NE $({\sigma^1}^\ast,{\sigma^2}^\ast)\in\mathcal{S}_{\Gamma}$, ${\sigma^1}^\ast$ (resp. ${\sigma^2}^\ast$) is a minimaximizer for player~$1$ (resp. player $2$); but the converse is not true. Also, for each player there is a unique maximinimizing strategy but it is not a Nash strategy.
For the sake of brevity, we have omitted the proofs of several lemmas and propositions. 

\subsection{Preliminary results}\label{preliminary_results}

Let us first present the following lemma which states that \defender does not choose a flow $\flow$ that contains a loop. 
\vspace{0.1cm}
\begin{lemma}
\textit{Any flow containing loops is not a best response for \defender.}
\end{lemma}
\vspace{0.1cm}
That is, \defender's all strategies that contain loops are strictly dominated, and they will not be in the support of a NE. Thus, $\Lambda$ in~\eqref{eq:pathstoedges} can be restricted to the set of $s-t$ paths. 

In the rest of this paper, we use the following notations: $\flow^0$ the action of not sending flow in the network, $\flow^{\ast}$ an optimal solution of $(\mathcal{P})$ that satisfies Assumption~\ref{big_assumption}, $\att^0$ the action of not attacking any edge of the network, and $\att^{min}$ the action that disrupts all the edges of a min-cut of the network. 
%



\begin{figure}[htbp]
\centering
\begin{tikzpicture}[x = 0.78cm,y=0.8cm]
\draw[thick][->] (0,0) -- (10,0) node[anchor=north] {$p_1$};
\draw	(0,-0.1) node[anchor=north] {$0$}
		(4,-0.1) node[anchor=north] {$\alpha$};

\draw[thick][->] (0,0) -- (0,4) node[anchor=east] {$p_2$};

\draw	(-0.1,2) node[anchor=east] {$1$};		
\draw[thick] (-0.07,2) -- (0.07,2);	

\draw[thick] (4,0) -- (4,4);
\draw[thick] (4,2) -- (10,2);

\draw	(2,2.4) node{{\textcolor{blue}{$\supp(\sigma^{1^*})=\{\flow^0\}$}}}
		(2,1.6) node{\textcolor{red}{$\supp(\sigma^{2^*})=\{\att^0\}$}}
		(7,1.4) node{\textcolor{blue}{$\supp(\sigma^{1^*})=\{\flow^{\ast}\}$}}
		(7,0.6) node{\textcolor{red}{$\supp(\sigma^{2^*})=\{\att^{0}\}$}}
		(7,3.4) node{\textcolor{blue}{$\supp(\sigma^{1^*})=\{\flow^0,\flow^{\ast}\}$}}
		(7,2.6) node{\textcolor{red}{$\supp(\sigma^{2^*})=\{\att^{0},\att^{min}\}$}};

\draw	(3.5,3.5) node {\textbf{I}\label{region 1}};
\draw	(9.5,1.5) node {\textbf{II}\label{region 2}};
\draw	(9.5,3.5) node {\textbf{III}\label{region 3}};

\end{tikzpicture}
\caption{Support of equilibrium strategies in Regions \textbf{I}-\textbf{III}.}
\label{fig:Plot Support}
\end{figure}
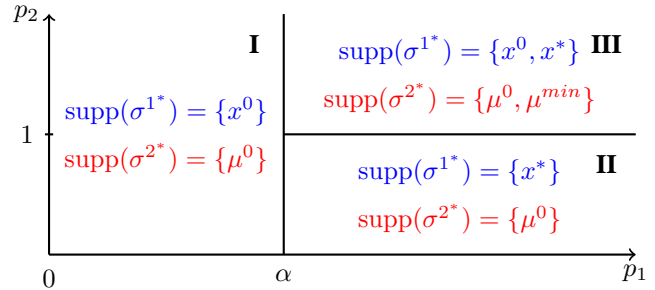

The following propositions~\ref{no flow}--\ref{sigma0} provide that, for a given $\alpha$, the game~$\Gamma$ admits qualitatively different  equilibria in regions $0 <p_1<\alpha$ and $p_2>0$ (Region~\textbf{I}), $p_1> \alpha$ and $0 < p_2< 1$ (Region~\textbf{II}), and $p_1> \alpha$ and $p_2>1$ (Region~\textbf{III}). These results are summarized by Fig.~\ref{fig:Plot Support}. 

The following result states that no flow and no attack is the unique NE of $\Gamma$ in Region~$\textbf{I}$:  
\vspace{0.1cm}
\begin{proposition}[Region~\textbf{I}]\label{prop:1}
\textit{If $0 < p_1 < \alpha$, then $\mathcal{S}_{\Gamma} = \{({\flow}^0,{\att}^0)\}$ with $u_1({\flow}^0,{\att}^0) = 0$ and $u_2({\flow}^0,{\att}^0) = 0$.}
\label{no flow}
\end{proposition}
\vspace{0.1cm}

Intuitively, when $0<p_1 <\alpha$, the marginal value of effective flow that reaches the destination node $t$ is less than the marginal transportation cost for every s-t path. Therefore, \defender will face negative utility if she sends flow through the network. Thus, in this case, her best response is not to route any flow. Since no flow is sent by \defender, \attacker's best response is not to attack, otherwise she faces the cost of attack without any value from lost flow. 

Next, for Region~\textbf{II}, we obtain that max-flow with minimum transportation cost and no attack is a pure NE: 
\vspace{0.1cm}
\begin{proposition}[Region~\textbf{II}]
\textit{If $p_1 > \alpha$ and $0<p_2 < 1$, then $\forall \flow^{\ast} \in \Omega, \ \{{\flow}^{\ast},{\att}^0\} \in \mathcal{S}_{\Gamma}$.
The corresponding payoffs are $u_1({\flow}^{\ast},{\att}^0) = \displaystyle (p_1-\alpha)\Theta$ and $u_2({\flow}^{\ast},{\att}^0) = 0$.}
\label{no attack}
\end{proposition}
\vspace{0.1cm}

This result can be explained as follows: on one hand, since \attacker's valuation of lost flow is small ($p_2<1$), for any attack, the utility gained from the lost flow is always lower than the cost of attack. Therefore, \attacker's best response is not to attack any edge. On the other hand, \defender's valuation of effective flow reaching $t$ is higher than the disutility it faces in transportation cost ($p_1>\alpha$). Since \attacker does not disrupt any edge, every flow sent through the network reaches $t$; thus, \defender's best response is to send a maximum flow. Among the different maximum flows, the max-flows with minimum transportation cost maximize \defender's equilibrium payoff. 

Not that if $p_1 = \alpha$ and $p_2 <1$, then both $(\flow^0,\att^0)$ and $(\flow^{\ast},\att^0)$ are NE. The equilibrium payoffs are still $(0,0)$.

The following proposition~\ref{sigma0} focuses on Region~\textbf{III}: 
\vspace{0.1cm}
\begin{proposition}[Region~\textbf{III}]\label{sigma0}
\textit{If $p_1 >\alpha$ and $p_2 >1$, then $\Gamma$ has no pure NE. Furthermore, $\exists \, \sigma_0 = (\sigma^1_0,\sigma^2_0) \in \mathcal{S}_{\Gamma}$ such that $U_1(\sigma^1_0,\sigma^2_0)=U_2(\sigma^1_0,\sigma^2_0)=0$, and $ \supp(\sigma^1_0) = \{{\flow}^0,{\flow}^{\ast}\}$ and $\supp(\sigma^2_0) = \{{\att}^0,{\att}^{min}\}$. The corresponding probabilities are given by:}
\begin{itemize}
\item $\displaystyle \sigma^1_{{\flow}^0} = 1 - \frac{1}{p_2}$, \quad $\displaystyle \sigma^1_{{\flow}^{\ast}} = \frac{1}{p_2}$\\
\item $\displaystyle \sigma^2_{{\att}^0} = \frac{\alpha}{p_1}$, \quad$\displaystyle \sigma^2_{{\att}^{min}} = 1 - \frac{\alpha}{p_1}.$\\
\end{itemize}
\end{proposition}
\vspace{0.1cm}

From this result, we can make several useful observations. First, in contrast to Prop.~\ref{no flow} and Prop.~\ref{no attack}, in Region~\textbf{III}, both players must randomize their actions in any equilibrium. Second, the equilibrium $(\sigma^1_0,\sigma^2_0)$, as defined in Prop.~\ref{sigma0}, can be obtained from a solution of problem $(\mathcal{P})$ and a min-cut of the graph~$\mathcal G$.\footnote{Recall that, in general, computing an equilibrium and enumerating all equilibria of general two-player non-zero sum games are computationally involved problems~\cite{AvisRosenberg,DBLP:journals/cacm/DaskalakisGP09}.} Third, $(\sigma^1_0,\sigma^2_0)$ has a  particularly simple structure, i.e., \defender either sends the whole max-flow with minimum transportation cost, or does not send any flow in the network. Similarly, \attacker either disrupts all the edges of a min-cut, or does not attack any edge of the network. 

Finally, Prop.~\ref{sigma0} is consistent with the game-theoretic intuition: \defender's equilibrium strategy $\sigma^1_0$ is characterized by the parameter $p_2$, and similarly, \attacker's equilibrium strategy $\sigma^2_0$ is characterized by the parameters $p_1$ and $\alpha$ (which is given and fixed under Assumption~\ref{big_assumption}). This intuition can be further explained as follows: as $p_2$ increases, $\sigma^1_{{\flow}^{\ast}}$ decreases while $\sigma^1_{{\flow}^0}$ increases. When \attacker's valuation of lost flow, $p_2$, is large, she has more incentive to attack, so any flow sent by \defender will be more likely to be lost. Thus, \defender chooses not to send any flow with higher probability than sending $\flow^{\ast}$. Likewise, as $p_1$ increases, $\sigma^2_{{\att}^{min}}$ increases while $\sigma^2_{{\att}^{0}}$ decreases. Again, when the marginal valuation of effective flow, $p_1$, is large, \defender will prefer to send as much flow as she can. Thus, \attacker will be more likely to gain by inducing an important loss to \defender by attacking a min-cut.
We illustrate Prop. \ref{no flow}, \ref{no attack} and \ref{sigma0} with an example.
\vspace{0.09cm}
\begin{example} 
Consider the graph in Fig.\ref{Example}. We can see that $\alpha = 3$, and that the max-flow with minimum transportation cost sends 1 unit of flow through $\{s,1,3,t\}$, $\{s,2,3,t\}$ and $\{s,2,4,t\}$ that induce a transportation cost equal to 3 each. Therefore, Assumption \ref{big_assumption} is satisfied. It is easy to see that the min-cut set is given by $\{(1,3),(2,3),(2,4)\}$.
\begin{figure}[htbp]
\centering
\begin{tikzpicture}[->,>=stealth',shorten >=1pt,auto,x=2cm, y=1.2cm,
  thick,main node/.style={circle,draw},flow_a/.style ={blue!100}]
\tikzstyle{edge} = [draw,thick,->]
\tikzstyle{cut} = [draw,very thick,-]
\tikzstyle{flow} = [draw,thick,->,blue!100]
	\node[main node] (s) at (0,1) {s};
	\node[main node] (1) at (1,2) {1};
	\node[main node] (2) at (1,0) {2};
	\node[main node] (3) at (2.5,2) {3};
	\node[main node] (4) at (2.5,0) {4};
	\node[main node] (t) at (3.5,1) {t};
	
	\path[edge]
	(2) edge node{\textcolor{red}{1},\textcolor{purple}{1}} (1)
	(4) edge node{\textcolor{red}{1},\textcolor{purple}{1}} (3)
	(s) edge node[above left]  {\textcolor{red}{2},\textcolor{purple}{1}} (1)
	(2) edge node{\textcolor{red}{1},\textcolor{purple}{1}} (3)
	(s) edge node[below left] {\textcolor{red}{3},\textcolor{purple}{1}} (2)
	(1) edge node  {\textcolor{red}{1},\textcolor{purple}{1}} (3)
	(2) edge node[below] {\textcolor{red}{1},\textcolor{purple}{1}} (4)	
	(3) edge node[above right]{\textcolor{red}{3},\textcolor{purple}{1}} (t)
	(4) edge[below right] node{\textcolor{red}{2},\textcolor{purple}{1}} (t);
	
\end{tikzpicture}
\caption{Example network. Edge capacities and transportation costs are denoted in red and purple labels respectively.}
\label{Example}
\end{figure}
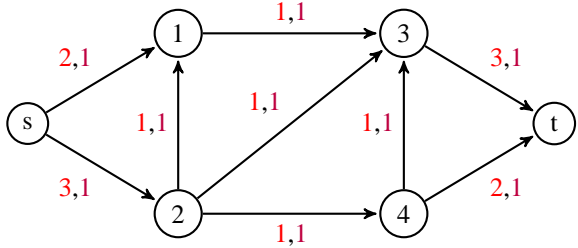

The NE described in Prop. \ref{no flow}, \ref{no attack} and \ref{sigma0} are illustrated in Fig.~\ref{NE1}.

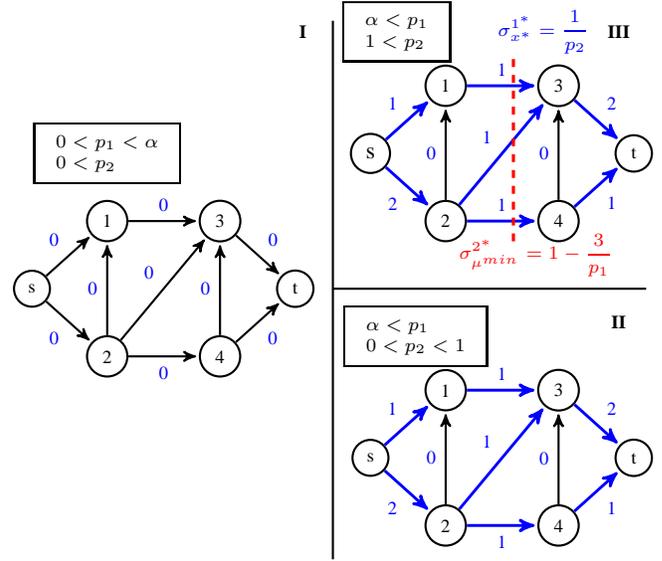
\begin{figure}[htbp]
\centering
\begin{tikzpicture}[->,>=stealth',shorten >=1pt,auto,x=1cm, y=0.9cm,
  thick,main node/.style={circle,draw,font=\small},flow_a/.style ={blue!100}]
\tikzstyle{edge} = [draw,thick,->]
\tikzstyle{cut} = [draw,very thick,-]
\tikzstyle{flow} = [draw,line width = 1.1pt,->,blue!100]
\tikzstyle{threshold} = [draw, thick,-]
	\draw[thick,threshold] (4,1) -- (8.2,1);
	\draw[thick,threshold] (4,-3) -- (4,5);

	\node[main node] (s_1) at (0,1) {\scriptsize s};
	\node[main node] (1_1) at (1,2) {\scriptsize1};
	\node[main node] (2_1) at (1,0) {\scriptsize2};
	\node[main node] (3_1) at (2.5,2) {\scriptsize3};
	\node[main node] (4_1) at (2.5,0) {\scriptsize4};
	\node[main node] (t_1) at (3.5,1) {\scriptsize t};

	\path[edge]
	(2_1) edge node{\scriptsize \textcolor{blue}{0}} (1_1)
	(4_1) edge node{\scriptsize \textcolor{blue}{0}} (3_1)
	(s_1) edge node[above left]  {\scriptsize \textcolor{blue}{0}} (1_1)
	(2_1) edge node{\scriptsize \textcolor{blue}{0}} (3_1)
	(s_1) edge node[below left] {\scriptsize \textcolor{blue}{0}} (2_1)
	(1_1) edge node  {\scriptsize\textcolor{blue}{0}} (3_1)
	(2_1) edge node[below] {\scriptsize\textcolor{blue}{0}} (4_1)	
	(3_1) edge node[above right]{\scriptsize\textcolor{blue}{0}} (t_1)
	(4_1) edge[below right] node{\scriptsize\textcolor{blue}{0}} (t_1);

\node[main node] (s_2) at (4.5,-1.5) {\scriptsize s};
	\node[main node] (1_2) at (5.5,-0.5) {\scriptsize1};
	\node[main node] (2_2) at (5.5,-2.5) {\scriptsize2};
	\node[main node] (3_2) at (7,-0.5) {\scriptsize3};
	\node[main node] (4_2) at (7,-2.5) {\scriptsize4};
	\node[main node] (t_2) at (8,-1.5) {\scriptsize t};
	
	\path[edge]
	(2_2) edge node{\scriptsize\textcolor{blue}{0}} (1_2)
	(4_2) edge node{\scriptsize\textcolor{blue}{0}} (3_2);
	
	\path[flow,every node/.style={black!100}]
	(s_2) edge node[above left]  {\scriptsize\textcolor{blue}{1}} (1_2)
	(2_2) edge node{\scriptsize\textcolor{blue}{1}} (3_2)
	(s_2) edge node[below left] {\scriptsize\textcolor{blue}{2}} (2_2)
	(1_2) edge node  {\scriptsize\textcolor{blue}{1}} (3_2)
	(2_2) edge node[below] {\scriptsize\textcolor{blue}{1}} (4_2)	
	(3_2) edge node[above right]{\scriptsize\textcolor{blue}{2}} (t_2)
	(4_2) edge[below right] node{\scriptsize\textcolor{blue}{1}} (t_2);

\node[main node] (s_3) at (4.5,3) {s};
	\node[main node] (1_3) at (5.5,4) {\scriptsize 1};
	\node[main node] (2_3) at (5.5,2) {\scriptsize 2};
	\node[main node] (3_3) at (7,4) {\scriptsize 3};
	\node[main node] (4_3) at (7,2) {\scriptsize 4};
	\node[main node] (t_3) at (8,3) {\scriptsize t};
	
	\path[edge]
	(2_3) edge node{\scriptsize\textcolor{blue}{0}} (1_3)
	(4_3) edge node{\scriptsize\textcolor{blue}{0}} (3_3);
	
	\path[flow,every node/.style={black!100}]
	(s_3) edge node[above left]  {\scriptsize\textcolor{blue}{1}} (1_3)
	(2_3) edge node{\scriptsize\textcolor{blue}{1}} (3_3)
	(s_3) edge node[below left] {\scriptsize\textcolor{blue}{2}} (2_3)
	(1_3) edge node  {\scriptsize\textcolor{blue}{1}} (3_3)
	(2_3) edge node {\scriptsize\textcolor{blue}{1}} (4_3)	
	(3_3) edge node[above right]{\scriptsize\textcolor{blue}{2}} (t_3)
	(4_3) edge[below right] node{\scriptsize\textcolor{blue}{1}} (t_3);
	
	\node at (6.7,1.5) {\textcolor{red}{{\scriptsize${\sigma_{\att^{min}}^{2^*}}=1 - \dfrac{3}{p_1}$}}};
	\node at (6.8,4.8) {\textcolor{blue}{{\scriptsize${\sigma_{\flow^{\ast}}^{1^*}} = \dfrac{1}{p_2}$}}};
  \draw[very thick, red, dashed,cut] 
    (6.4,1.7).. controls +(right:0cm) and +(down:0cm) ..(6.4,4.4);

	\node at (1,3) {\fbox{\scriptsize$\begin{array}{l}0 < p_1 < \alpha \\0 < p_2\end{array}$}};
	\node at (5.1,0.3) {\fbox{\scriptsize$\begin{array}{l}\alpha < p_1 \\0 < p_2 < 1\end{array}$}};
	\node at (4.85,4.8) {\fbox{\scriptsize$\begin{array}{l}\alpha < p_1  \\1 < p_2\end{array}$}};
	\node at (3.6,4.8){\scriptsize\textbf{I}};
	\node at (7.8,0.5) {\scriptsize\textbf{II}};
	\node at (7.8,4.8){\scriptsize\textbf{III}};

\end{tikzpicture}
\caption{NE described in Prop. \ref{no flow}, \ref{no attack} and \ref{sigma0}.}
\label{NE1}
\end{figure}
\normalsize

\end{example}
\vspace{0.1cm}
%

\subsection{Main Theorem}\label{ss:mainthm}
We now present our main result which identifies several useful properties satisfied by \emph{all} NE of $\Gamma$ in Region~\textbf{III} and under Assumption~\ref{big_assumption}. 

\vspace{0.1cm}
\begin{theorem}\label{thm:mainresult}
\textit{If $p_1> \alpha$ and $p_2>1$, and under Assumption \ref{big_assumption}, then for any $\sigma^* \in \mathcal{S}_{\Gamma}$:}
\begin{enumerate}
\item \textit{Both players' equilibrium payoffs are equal to $0$, i.e.:}
\begin{align}
U_1({\sigma^1}^*,{\sigma^2}^*) &\equiv 0 \label{u1=0_st}\\
U_2({\sigma^1}^*,{\sigma^2}^*) &\equiv 0 \label{u2=0_st}
\end{align}

\item \textit{The expected amount of initial flow is given by:}
\begin{align}
\Expone[*]{\Value{\flow}} \equiv  \frac{1}{p_2} \Theta 
\label{eq_exp_flow_st}
\end{align}

\textit{and the expected transportation cost is given by:}
\begin{align}
\Expone[*]{\Costf{\flow}} \equiv \frac{\alpha}{p_2} \Theta 
\label{eq_exp_transp_st}
\end{align}

\item \textit{The expected cost of attack is given by:}
\begin{align}
 \Exptwo[*]{\Cost{\att}}\equiv \left(1 - \frac{\alpha}{p_1}\right)\Theta 
 \label{eq_exp_att_st}
\end{align}

\item \textit{The expected amount of effective flow is given by:}
\begin{align}
\Expboth{*}{*}{\Eff{\flow}{\att}}\equiv \frac{\alpha}{p_1p_2}\Theta.
\label{exp_net_flow_st}
\end{align}

%
%
\end{enumerate}
\end{theorem}
\vspace{0.1cm}
We refer the reader to Appendix for the proof. From Thm.~\ref{thm:mainresult} we observe that although $\mathcal{S}_{\Gamma}$ is uncountable (following the convexity of $\mathcal{S}_{\Gamma}$), the equilibrium values of common characteristics, such as expected amount of initial and effective flow, and expected transportation cost to \defender and the expected cost of attack to \attacker, can be computed in closed-form using the parameters of the game and using the optimal value $\Theta$ of the max-flow problem. The payoffs of both players are constant for any NE (ref.~\eqref{u1=0_st} and \eqref{u2=0_st}). 
%
Below we further explain the implications of Thm.~\ref{thm:mainresult}. 

First, following~\eqref{eq_exp_flow_st}, the expected amount of initial flow is always equal to some fraction of the amount of max-flow, and this expectation decreases with $p_2$. 
Likewise, following~\eqref{eq_exp_att_st}, the expected cost of attack at any NE is always equal to some fraction of the cost of attacking a min-cut, and this expectation increases with $p_1$.




Secondly, following \eqref{eq_exp_transp_st}, the expected cost of transportation is always equal to some fraction of the cost of transporting $\flow^{\ast}$ (indeed, $\Costf{\flow^{\ast}} = \alpha \Theta$ by Assumption~\ref{big_assumption}). Notice that while $\alpha$ always governs the equilibrium strategy of \attacker, it also affects \defender's equilibrium transportation cost.

Third, following \eqref{exp_net_flow_st}, the expected amount of effective flow is always equal to some fraction of the amount of max-flow, and since $\frac{\alpha}{p_1 p_2} < \frac{1}{p_2}$ this flow is always smaller than the expected amount of initial flow. The expected amount of effective flow decreases when $p_1$ and/or $p_2$ increase. This is surprising because one might expect that if the marginal value of effective flow increases, its expectation would increase as well. This result can be explained by noting that when $p_1$ increases, the disruption caused by  \attacker increases, so there is more lost flow and the expected effective flow decreases.

Thm.~\ref{thm:mainresult} also enables estimation of the performance metrics such as the expected amount of lost flow and the yield of \defender in any NE. We define the yield as the ratio of the expected amount of flow that reaches $t$ (effective flow) and the expected amount of flow that is sent through $s$ (initial flow). We have the following corollary: 
\vspace{0.1cm}
\begin{corollary}
\textit{The expected amount of lost flow is given by:}
\begin{align}
\Expboth{*}{*}{\Loss{\flow}{\att}}\equiv \frac{1}{p_2}\left(1 -\frac{\alpha}{p_1}\right)\Theta,
\label{eq_exp_loss_st}
\end{align}
\textit{and the expected yield is given by:}
\begin{align}
\frac{\Expboth{*}{*}{\Eff{\flow}{\att}}}{\Expone[*]{\Value{\flow}}} \equiv \frac{\alpha}{p_1}.
\label{eq_exp_yield_st}
\end{align}
\end{corollary}
\vspace{0.1cm}

From~\eqref{eq_exp_loss_st}, we note that the expected amount of lost flow is always equal to some fraction of the amount of max-flow. The corresponding coefficient increases with $p_1$, because when $p_1$ is large, \defender sends more flow and \attacker disrupts more edges. However, the coefficient decreases when $p_2$ increases, mainly because when $p_2$ is large, \attacker causes more disruption and \defender sends less flow in the network. Interestingly, we obtain that the expected yield is a constant for all NE, and it only depends on $\alpha$ and $p_1$;  neither $p_2$ nor the maximum amount of flow $\Theta$ affect the yield. 


Finally, it is easy to check that all of these properties are satisfied by $(\sigma^1_0,\sigma^2_0)$ defined in Prop.~\ref{sigma0}.

\subsection{Additional properties}\label{ss:additional}

We now discuss additional properties satisfied by the NE of $\Gamma$. 
The following result relates the NE of $\Gamma$ with min-cuts of the network~$\mathcal G$. 

\vspace{0.1cm}
\begin{proposition}
$\forall ({\sigma^1}^*,{\sigma^2}^*) \in \mathcal{S}_{\Gamma}, \ \forall \att \in \supp({\sigma^2}^*)$:
\begin{align*}
&\Cost{\att} \leq \Cost{\att^{min}} = \Theta,\\
&\forall (i,j) \in \mathcal{E}, \ \att_{ij} = 1 \Longrightarrow \forall \flow^{\ast} \in \Omega, \ \flow^{\ast}_{ij} = c_{ij}.
\end{align*}
\end{proposition}
\vspace{0.1cm}

From this result, we obtain that, in equilibrium, there will be no attack that requires a cost of attack larger than the cost of attacking a min-cut. In addition, for any NE, an edge is disrupted with positive probability only if it is saturated by every maximum flow with minimum transportation cost. In other words, if there exists at least one maximum flow with minimum transportation cost that does not saturate an edge, then this edge will never be disrupted at equilibrium.

\vspace{0.1cm}
\begin{proposition}
\textit{Consider a min-cut $E(\{S,T\})$, then:}
\begin{align*}
\forall ({\sigma^1}^*,{\sigma^2}^*) \in \mathcal{S}_{\Gamma}, \ \forall (i,j) \in E(\{S,T\}), \ \mathbb{E}_{{\sigma}^*}[\flow_{ij}] = \frac{c_{ij}}{p_2}, 
\end{align*}

%
%

\textit{Furthermore, for any NE whose support is based on attacks that only disrupt edges of $E(\{S,T\})$ we have:}
$$\forall (i,j) \in E(\{S,T\}), \ \mathbb{P}\left((i,j) \ \text{is disrupted}\right) = 1 - \frac{\alpha}{p_1}.$$
\end{proposition}
\vspace{0.1cm}

This proposition tells us that at any NE, the expected amount of flow that goes through any edge of a min-cut is always equal to a fraction of its capacity. In addition, if \attacker's equilibrium strategy only disrupts edges of one min-cut, then the probability with which an edge is disrupted is constant for all the edges of that min-cut, irrespective of the capacities of these edges.
\vspace{0.1cm}
\begin{corollary}$\forall ({\sigma^1}^*,{\sigma^2}^*) \in \mathcal{S}_{\Gamma}, \ \forall \ \textit{min-cut} \ E(\{S,T\})$, 

$\forall (i,j) \in E(\{S,T\}): \ \exists \, \flow \in \supp({\sigma^1}^*) \ | \ \flow_{ij} >0$,
\label{flows}
\end{corollary}
\vspace{0.1cm}

This corollary means that for any NE and for any edge of a min-cut, there exists a flow chosen with non-zero probability that passes through it.

All of these properties enable us to derive the following bounds which are tight thanks to Prop.~\ref{sigma0}:

\vspace{0.1cm}
\begin{proposition}
\textit{Consider $({\sigma^1}^*,{\sigma^2}^*) \in \mathcal{S}_{\Gamma}$. Then we have the following bounds: }
\begin{itemize}
\item \textit{If $\flow^0 \in \supp({\sigma^1}^*)$, then
$\sigma_{\flow^0}^{1^*} \leq 1 - \dfrac{1}{p_2}$,}

\item\textit{If $\flow^{\ast} \in \supp({\sigma^1}^*)$, then ${\sigma_{\flow^{\ast}}^{1^*}} \leq \dfrac{1}{p_2}$,}

\item \textit{If $\att^{min} \in \supp({\sigma^2}^*)$, then ${\sigma_{\att^{min}}^{2^*}} \leq 1 - \dfrac{\alpha}{p_1}$,}

\item \textit{If $\att^0 \in \supp({\sigma^2}^*)$, then ${\sigma_{\att^{0}}^{2^*}} \leq \dfrac{\alpha}{p_1}$.}
\end{itemize}
\end{proposition}
\vspace{0.1cm}

This proposition gives bounds on the probability with which $\flow^0, \flow^{\ast}, \att^0, \att^{min}$ are chosen when they are in the support of a NE. From these upper bounds, we see that when $p_2$ is close to $1$, the probability with which $\flow^0$ can be chosen becomes very small. On the contrary, it is when $p_2$ is large that $\flow^{\ast}$ can be chosen with small probability. Similarly, when $p_1$ is close to $\alpha$, $\att^{min}$ can be chosen only with small probability, and when $p_1$ is large, $\att^0$ can be chosen only with small probability.

Recall the classical result in SCG that maximinimizers are also NE. In the game~$\Gamma$, ${\flow}^0$ is a \emph{maximinimizer} for \defender and ${\att}^0$ is a \emph{maximinimizer} for \attacker. However, we already know that if $p_1 >\alpha$ and $p_2 >1$, there is no pure NE. Hence, the maximinimizing strategies do not give NE, i.e., $({\flow}^0,{\att}^0) \notin \mathcal{S}_{\Gamma}$. Still, we have the following result:

\vspace{0.1cm}
\begin{proposition}
\textit{Each player's payoffs for both maximinimizing and minimaximizing strategies are equal to the payoff at NE, i.e., } 
\begin{align*}
	 \max_{\sigma^1} \min_{\sigma^2} U_1(\sigma^1,\sigma^2) = 0 = \min_{\sigma^2}\max_{\sigma^1}U_1(\sigma^1,\sigma^2)\\
\max_{\sigma^2} \min_{\sigma^1} U_2(\sigma^1,\sigma^2) = 0 = \min_{\sigma^1}\max_{\sigma^2}U_2(\sigma^1,\sigma^2)
\end{align*}
\textit{Furthermore, the set of minimaximizers is a superset of $\mathcal{S}_{\Gamma}$, i.e., any NE is a minimaximizer.}
\end{proposition}
\vspace{0.1cm}

\section{Relaxing Assumption 1}\label{sec:relax}
We now discuss an example that does not satisfy Assumption \ref{big_assumption}. We consider the network given in Fig. \ref{pathological}:
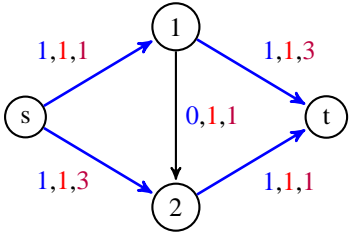
\begin{figure}[htbp]
\centering
\begin{tikzpicture}[->,>=stealth',shorten >=1pt,auto,x=2cm, y=1.2cm,
  thick,main node/.style={circle,draw},flow_a/.style ={blue!100}]
\tikzstyle{edge} = [draw,thick,->]
\tikzstyle{cut} = [draw,very thick,-]
\tikzstyle{flow} = [draw,line width = 1pt,->,blue!100]
	\node[main node] (s) at (0,1) {s};
	\node[main node] (1) at (1,2) {1};
	\node[main node] (2) at (1,0) {2};
	\node[main node] (t) at (2,1) {t};
	
	\path[edge]
	(1) edge node{\textcolor{blue}{0},\textcolor{red}{1},\textcolor{purple}{1}} (2);
	
	\path[flow,every node/.style={black!100}]
	(s) edge node[above left]  {\textcolor{blue}{1},\textcolor{red}{1},\textcolor{purple}{1}} (1)
	(s) edge node[below left] {\textcolor{blue}{1},\textcolor{red}{1},\textcolor{purple}{3}} (2)
	(1) edge node[above right]{\textcolor{blue}{1},\textcolor{red}{1},\textcolor{purple}{3}} (t)
	(2) edge[below right] node{\textcolor{blue}{1},\textcolor{red}{1},\textcolor{purple}{1}} (t);
	
\end{tikzpicture}
\caption{Network that does not satisfy Assumption \ref{big_assumption}.}
\label{pathological}
\end{figure}

The unique max-flow with minimum transportation cost is drawn in blue in Fig. \ref{pathological}. It sends one unit of flow through path $\{s,1,t\}$ and one unit through path $\{s,2,t\}$. They both induce a marginal transportation cost equal to 4. However, $\{s,1,2,t\}$ has a marginal transportation cost equal to 3, so Assumption \ref{big_assumption} does not hold.

Now, we suppose that $3 < p_1 < 4$ and $p_2 >1$ then we can prove that $({\sigma^1}^*,{\sigma^2}^*)$ defined by ${\sigma^{1^*}_{\flow^0}} = 1 - \dfrac{1}{p_2}$, ${\sigma^{1^*}_{\flow^1}} = \dfrac{1}{p_2}$, ${\sigma^{2^*}_{\att^0}} = \dfrac{3}{p_1}$, ${\sigma^{2^*}_{\att^1}} = 1 -\dfrac{3}{p_1}$ is a NE ($\flow^1$ sends one unit of flow through path $\{s,1,2,t\}$ and $\att^1$ disrupts edge $(1,2)$).

We can show that, in this case, the NE we presented earlier cannot be found anymore, and the one we found is not supported by any max-flow with minimum transportation cost and any min-cut.

\section{Appendix: Proof of Theorem 1}\label{sec:appendix}
We will need the following lemma in the proof of the main theorem: 
\vspace{0.1cm}
\begin{lemma}
\begin{align}
\forall ({\sigma^1}^*,{\sigma^2}^*) \in \mathcal{S}_{\Gamma}, \ \Exptwo[*]{\Effp{\flow^{\ast}}{\att}} = \Theta -\Exptwo[*]{\Cost{\att}}
\label{nice_eq}
\end{align}
\label{intermediate}
\end{lemma}
\vspace{0.1cm}
\textit{Proof of Lemma \ref{intermediate}:}
We can find a link between the expected payoffs that we can write in two different ways:
\begin{align}
U_1(\sigma^1,\sigma^2) =&p_1\Expone{\Value{\flow}} -\Expone{\Costf{\flow}} \nonumber\\
&- \frac{p_1}{p_2}  \Exptwo{\Cost{\att}} - \frac{p_1}{p_2}U_2(\sigma^1,\sigma^2) \label{link1}\\
U_2(\sigma^1,\sigma^2)  = &-  \Exptwo{\Cost{\att}} +p_2 \Expone{\Value{\flow}}\nonumber \\
&-\frac{p_2}{p_1}\Expone{\Costf{\flow}} - \frac{p_2}{p_1}U_1(\sigma^1,\sigma^2) \label{link2}
\end{align}
Let $\sigma^* = ({\sigma^1}^*,{\sigma^2}^*) \in \mathcal{S}_{\Gamma}$. Since $(\sigma^1_0,\sigma^2_0) \in \mathcal{S}_{\Gamma}$ (Prop. \ref{sigma0}), (\ref{best2}) and (\ref{payoff2}) give us:
\begin{align*}
0 = U_2(\sigma^1_0,\sigma^2_0)  &\geq U_2(\sigma^1_0,{\sigma^2}^*) \\
&= \Theta - \Exptwo[*]{\Effp{\flow^{\ast}}{\att}} - \Exptwo[*]{\Cost{\att}}
\end{align*}
So we get the first inequality:
\begin{align}
\Exptwo[*]{\Effp{\flow^{\ast}}{\att}} \geq \Theta - \Exptwo[*]{\Cost{\att}} 
\label{1st_part_ineq}
\end{align}
Now, since $({\sigma^1}^*,{\sigma^2}^*) \in \mathcal{S}_{\Gamma}$, (\ref{best1}), (\ref{payoff1}) and Assumption 1 give:
\begin{align}
U_1({\sigma^1}^*,{\sigma^2}^*) &\geq \frac{p_1}{p_2}\Exptwo[*]{\Effp{\flow^{\ast}}{\att}} - \frac{\alpha}{p_2} \Value{\flow^{\ast}}\label{u1down}
\end{align}
If we combine (\ref{link1}), (\ref{payoff2}) and (\ref{best2}), using $\sigma^2_0$, we get:
\begin{align}
U_1({\sigma^1}^*,{\sigma^2}^*) \leq &\frac{p_1}{p_2} \left(  \Cost{\att^{min}} - \Exptwo[*]{\Cost{\att}}\right)\nonumber \\
&- \frac{\alpha}{p_2}\Cost{\att^{min}}
\label{u1up}
\end{align}
%
By combining (\ref{u1down}) and (\ref{u1up}), and using the max-flow min-cut theorem, we obtain:
\begin{align}
\Exptwo[*]{\Effp{\flow^{\ast}}{\att}} \leq \Theta -\Exptwo[*]{\Cost{\att}}
\label{2nd_part_ineq}
\end{align}
Equations (\ref{1st_part_ineq}) and (\ref{2nd_part_ineq}) lead to:
\begin{align*}
\Exptwo[*]{\Effp{\flow^{\ast}}{\att}} = \Theta -\Exptwo[*]{\Cost{\att}}
\end{align*}
\hfill \QED

We are now ready to prove the main result.
\vspace{0.1cm}

\textit{Proof of Theorem \ref{thm:mainresult}:}
Let $\sigma^* = ({\sigma^1}^*,{\sigma^2}^*) \in \mathcal{S}_{\Gamma}$. Let us first show 3), but we will need a few equations before. 
First, let's prove that $U_1({\sigma^1}^*,{\sigma^2}^*) \geq 0$. Equation (\ref{best1}) gives:
\begin{align}
U_1({\sigma^1}^*,{\sigma^2}^*) \geq U_1(\flow^0,{\sigma^2}^*) = 0
\label{u1>0}
\end{align}
If we combine (\ref{u1>0}) and (\ref{u1up}), we obtain:
\begin{align}
 \Exptwo[*]{\Cost{\att}} \leq \left(1 - \frac{\alpha}{p_1}\right) \Cost{\att^{min}}
 \label{1st_ineq_exp_att}
\end{align}
In order to get the reverse inequality, let us consider the strategy $\sigma^1_{\epsilon}$ defined by $\sigma^1_{{\flow}^{\ast}}  = \displaystyle\frac{1+\epsilon}{p_2}$ and $\sigma^1_{{\flow}^{0}}  = \displaystyle 1 - \frac{1+\epsilon}{p_2}$. For an $\epsilon$ small enough, (\ref{best1}), (\ref{payoff1}) and (\ref{nice_eq}) give us:
\begin{align}
U_1({\sigma^1}^*,{\sigma^2}^*)  \geq &\frac{p_1(1 + \epsilon)}{p_2}\left( \Value{\flow^{\ast}} -\Exptwo[*]{\Cost{\att}} \right) \nonumber\\
&- \frac{\alpha(1 + \epsilon)}{p_2}\Value{\flow^{\ast}}
\label{u1downbis}
\end{align}
%
%
We just need to combine (\ref{u1up}) and (\ref{u1downbis}) in order to get:
\begin{align}
 \Exptwo[*]{\Cost{\att}} \geq \left(1 - \frac{\alpha}{p_1}\right)\Value{\flow^{\ast}}
 \label{2nd_ineq_exp_att}
\end{align}
Equations (\ref{1st_ineq_exp_att}), (\ref{2nd_ineq_exp_att}) and the max-flow min-cut theorem give us 3):
\begin{align}
 \Exptwo[*]{\Cost{\att}} = \left(1 - \frac{\alpha}{p_1}\right) \Cost{\att^{min}} =  \left(1 - \frac{\alpha}{p_1}\right)\Theta
 \label{eq_exp_att}
\end{align}

We can now use this equation to prove that \defender's payoff is equal to 0 at equilibrium:
combining (\ref{eq_exp_att}) and (\ref{u1up}) lead to:
\begin{align}
U_1({\sigma^1}^*,{\sigma^2}^*) \leq 0
\label{u1<0}
\end{align}
Therefore, (\ref{u1<0}) and (\ref{u1>0}) give:
\begin{align}
U_1({\sigma^1}^*,{\sigma^2}^*) =  0
\label{u1=0}
\end{align}

Let's now show (\ref{eq_exp_flow_st}). Similarly, let's first prove that $U_2({\sigma^1}^*,{\sigma^2}^*) \geq 0$ using (\ref{best2}):
\begin{align}
U_2({\sigma^1}^*,{\sigma^2}^*)\geq U_2({\sigma^1}^*,\att^0) =0
\label{u2>0}
\end{align}
Then, (\ref{link2}), (\ref{u1=0}) and (\ref{eq_exp_att}) give:
\begin{align}
U_2({\sigma^1}^*,{\sigma^2}^*)  \leq \left(1 - \frac{\alpha}{p_1}\right) \left(p_2\Expone[*]{\Value{\flow}} - \Cost{\att^{min}}\right)
\label{u2up}
\end{align}
%
%
If we combine (\ref{u2up}) and (\ref{u2>0}), we obtain:
\begin{align}
\Expone[*]{\Value{\flow}} \geq \frac{1}{p_2}\Cost{\att^{min}}
\label{1st_ineq_exp_flow}
\end{align}
In order to get the reverse inequality, let us consider the strategy $\sigma^2_{\epsilon}$ defined by $\sigma^2_{{\att}^0} =\displaystyle \frac{\alpha - \epsilon}{p_1}$ and $\sigma^2_{{\att}^{min}} =\displaystyle  1 - \frac{\alpha - \epsilon}{p_1}$. For an $\epsilon$ small enough, (\ref{link1}), (\ref{best2}), (\ref{payoff2}) and (\ref{eq_exp_att}) give:
\begin{align}
U_1({\sigma^1}^*,{\sigma^2}^*) \leq \frac{\epsilon}{p_2}\Cost{\att^{min}} - \epsilon  \Expone[*]{\Value{\flow}}
%
%
%
 \label{epsilon2}
 \end{align}
%
%
 Equations (\ref{epsilon2}) and (\ref{u1>0}) give us:
%
\begin{align}
\Expone[*]{\Value{\flow}} \leq \frac{1}{p_2}\Cost{\att^{min}}
\label{2nd_ineq_exp_flow}
\end{align}
Equations (\ref{1st_ineq_exp_flow}), (\ref{2nd_ineq_exp_flow}) and the max-flow min-cut theorem give us:
\begin{align}
\Expone[*]{\Value{\flow}} = \frac{1}{p_2} \Value{\flow^{\ast}} = \frac{1}{p_2} \Theta
\label{eq_exp_flow}
\end{align}

Likewise, if we combine (\ref{eq_exp_flow}) and (\ref{u2up}), we get:
\begin{align}
U_2({\sigma^1}^*,{\sigma^2}^*) \leq 0
\label{u2<0}
\end{align}
Equations (\ref{u2<0}) and (\ref{u2>0}) give us:
\begin{align}
U_2({\sigma^1}^*,{\sigma^1}^*) = 0
\label{u2=0}
\end{align}
Thus 1) holds.
 
Now, if we combine (\ref{payoff2}), (\ref{eq_exp_flow}), (\ref{eq_exp_att}) and (\ref{u2=0}), we show 4):
\begin{align}
\Expboth{*}{*}{\Eff{\flow}{\att}}=\frac{\alpha}{p_1p_2}\Theta
\label{exp_net_flow}
\end{align}

Lastly, if we combine (\ref{payoff1}), (\ref{exp_net_flow}) and (\ref{u1=0}), we can show 2):
\begin{align}
\Expone[*]{\Costf{\flow}} \frac{\alpha}{p_2}\Theta
\label{eq_exp_cost}
\end{align}
\hfill
\QED

%

\addtolength{\textheight}{-12cm}   



%

\section*{Acknowledgments}

This work was supported in part by FORCES (Foundations
Of Resilient CybEr-Physical Systems), which
receives support from the National Science Foundation
(NSF award numbers CNS-1238959, CNS-1238962,
CNS-1239054, CNS-1239166), NSF CAREER award CNS-1453126, and the AFRL LABLET - Science of Secure and Resilient Cyber-Physical Systems (Contract ID: FA8750-14-2-0180, SUB 2784-018400).


\bibliographystyle{IEEEtran}
\bibliography{routing_games.bib}

\end{document}